\documentclass[11pt]{amsart}

\usepackage{epsfig}
\usepackage{verbatim}
\usepackage{amssymb}
\usepackage{amsfonts}
\usepackage{amsbsy}
\usepackage{graphicx}
\usepackage{color,subfigure,graphicx,longtable,citesort}

\newtheorem{Def}{Definition}[section]

\begin{document}

\title{Stability of the World Trade Web over time -- an extinction analysis}
\author{Nick Foti$^\dagger$}
\thanks{$^\dagger$Department of Computer Science, Dartmouth College, Hanover, NH 03755}
\author{Scott Pauls$^*$}
\thanks{$^*$Department of Mathematics, Dartmouth College, Hanover, NH 03755}
\author{Daniel N. Rockmore$^{\dagger,*,\#}$}
\thanks{\small $^{\#}$The Santa Fe Institute, Santa Fe, NM 87501}
%\footnote{}{\small $^\dagger$Department of Computer Science, Dartmouth College, Hanover, NH 03755}
%\affiliation{\small $^*$Department of Mathematics, Dartmouth College, Hanover, NH 03755}
%\affiliation{\small $^{\#}$The Santa Fe Institute, Santa Fe, NM 87501}
\date{\today}

\begin{abstract}
The World Trade Web (WTW) is a weighted network whose nodes correspond to countries with edge weights reflecting the value of imports and/or exports between countries. In this paper we introduce to this macroeconomic system the notion of {\em extinction analysis}, a technique often used in the analysis of ecosystems, for the purposes of investigating  the robustness of this network. In particular, we subject the WTW to a principled set of {\em in silico} ``knockout experiments," akin to those carried out in the investigation of food webs, but suitably adapted to this macroeconomic network. Broadly, our experiments show that over time the WTW moves to a ``robust yet fragile" configuration where it is robust to random failures but fragile under targeted attack.  This change in stability is highly correlated with the connectance (edge density) of the network.  Moreover, there is evidence of a sharp change in the structure of the network in the 1960s and 1970s, where  most measures of robustness rapidly increase before resuming a declining trend.  We interpret these results in the context in the post-World War II move towards globalization.  Globalization coincides with the sharp increase in robustness but also with a rise in those measures (e.g., connectance and trade imbalances) which correlate with decreases in robustness.  The peak of robustness is reached after the onset of globalization policy but before the negative impacts are substantial. These analyses depend on a simple model of dynamics that rebalances the trade flow upon network perturbation, the most dramatic of which is node deletion. More subtle and textured forms of perturbation lead to the definition of other measures of node importance as well as vulnerability. We anticipate that experiments and measures like these can play an important role in the evaluation of the stability of economic systems.
\end{abstract}

\maketitle

\section{Introduction}
In this paper we introduce new methods to articulate measures of robustness in economic networks.  In these, and
other living complex systems, researchers interested in studying   robustness generally do not have the luxury of performing {\em in vivo} experiments to test hypotheses. Consider the example of an ecosystem:  we cannot remove species in order to explore the impact of their sudden extinction. Similarly,  we cannot remove or disable components of an economy to discover the downstream effects.  However, with a sensible model we do have the ability to perform  simulations, i.e., {\em in silico} experiments designed to shed light on the interdependence of the system's components, especially as concerns  the understanding of those factors critical to  robustness of the system to sudden component failure.  
This is the sort of methodology that we aim to bring to economic systems, and in particular, to the World Trade Web (WTW). The WTW is an economic network summarizing international trade. Nodes represent countries and country $B$ links to country $A$ with a weight given by the value (say in US dollars) that  country $A$ receives from country $B$ in exports. With the rise of network analysis the WTW has received a good deal of attention and analysis (see e.g., \cite{F2,fagiolo-2007,serrano-2003-68}).

Previous network analyses of the WTW have provided a wealth of initial information regarding the interaction of the network structure and the functional architecture of the WTW.  Our goal is to extend this analysis and to craft measures of robustness or stability that take into account the intertwined structure and aspects of dynamics.  Our core methodology is that of a {\em knockout experiment}, where stability is tested via perturbation of the network by removal or degradation of some aspect of the structure, which in turn initiates some dynamic reorganization of the network.  We then observe the system as it returns to equilibrium and record the effect.

Knockout experiments  have already been performed in a variety of contexts, including the World Wide Web (WWW)~\cite{Albert00}, metabolic networks \cite{Jeong2000}, protein networks \cite{Jeong2001},  and, of particular relevance to this paper, in the form of {\em extinction analyses} conducted on food web models of ecosystems (see e.g., Section 4.6 of \cite{dunne2006} and the many references therein as well as  \cite{Allesina2009}). %Our main contribution is the transportation of the methods of extinction analysis in food webs [cite] to the setting of economically derived networks.
  Food webs are network models of ecological systems (see \cite{dunne2006} and references therein) where the nodes  represent species in the ecosystem and a directed link is placed to node $A$ from node $B$ if species $A$ eats species $B$ (thereby encoding the transfer of resource from $B$ to $A$).
  %Appropriate dynamics, derived from field data, used to simulate reasonable population fluctuations can also be incorporated to add dynamics \cite{}.
In this setting, stability and robustness are analyzed via simulated extinction studies in which certain species or groups of species are removed from the ecosystem.  ``Evolution" then takes place according to the simulated extinction dynamics. In its simplest (but most widely used) form, species removal entails the removal of all links to and from the corresponding node and further extinctions occur as species lose all of their resources (prey). The  system either reaches a new stable configuration (possibly after one or more consequent extinctions) or collapses entirely.  Such studies address questions of robustness by measures of the existence and extent of extinction cascades induced by the knockout of particular species.

The analogies that might be drawn between food webs and economies have a long history (see e.g., \cite{jacobs}) and are of great interest. In this paper we bring the idea of extinction analyses and related robustness analysis to the  WTW.  Herein we provide three types of knockout experiments to that end.  First, we consider an extinction analysis similar to that used in the study of food webs where countries are sequentially removed from the trade web and their impacts analyzed.  We then consider a variant in which rather than removing countries, we instead perturb their import/export profile.  Lastly, we consider deletion of edges in the network -- extinction of trade relationships -- and analyze their impact.  Each of these studies depends on a simple model of ``resource rebalancing" that drives a dynamics of WTW reorganization in response to the initial shock (see Section 2). Thus, we emphasize that  the results of these experiments must be interpreted as consequences of both the network structure of the WTW as well as the evolution dynamics we place on that structure.  In particular, more or less refined models of evolution dynamics may yield more or less textured results.  We view the dynamics described herein as a parsimonious first model that we hope can serve as either an inspiration or a foundation for subsequent analyses.

The integration of new network-related  methodologies to economic situations is of great current interest \cite{F1}. The coupling of dynamics with network thinking bears some relation to the new contagion analysis that is now finding some purchase in the economics literature (see e.g., \cite{AAE,AGKMP,BST}).
%.  In this a propagation dynamics is assumed on a network structure and their combined effect on the spread of a contagion through the network is investigated.  Dynamics models range from relatively simple diffusion or Markovian dynamics \cite{} to more general  network investigations (see, for example, \cite{sw1,sw2,Watts:Nature:1998,dw2004,Albert00}).  In our case, as with many economic systems, these types of simple dynamics are too coarse or otherwise not appropriate.  To compensate for this, we use dynamics that mirror aspects of the underlying economic mechanisms.  As such, our work is more similar to adapted contagion studies of economic systems (see e.g., \cite{AAE,AGKMP,BST}).

In our first analysis we introduce the notion of  {\em maximal extinction analysis} (MEA)  for the study of the WTW. The mechanism of MEA  is analogous to the kinds of knockout experiments performed on food webs, but differs in the formulation of the extinction dynamics (i.e., in the hypotheses and mechanisms
that dictate the consequences of node removal for the reduced network). Our methodology
is described in detail in the next section. The output of the accompanying simulation is a measurement
of the {\em extinction power} of any given country over any other. Roughly, the extinction power
of country $A $ over country $B$ is the proportion of $B$'s economic activity (as measured by
export value) disrupted by country $A$'s extinction (as well as any extinctions which are a
consequence of $A$'s). We view this as a measure of ``economic centrality."

To create an aggregate statistic, we define the {\em robustness} of a trade network analogously to that of a food web, namely the proportion of the total income in the network needed to be destroyed via iterative worst case node deletion and consequent extinctions to reach the loss of at least 50\% of the income in the network (the food web definition looks instead at the minimal fraction of species that must be removed to result in at least a 50\% reduction in the number of species).  This notion of robustness gives a measure of stability of the trade network --  it measures the extent to which countries can replace lost imports and lost demand for exports, given a shock in the system.  In tracking robustness over this period 1870--2006, we see two basic results.
\begin{enumerate}
\item There are two sharply different periods, around World War I (1914--1919) and World War II (1939--1948), where the robustness is dramatically higher than at any other time.  In both of these cases, the data sets have substantial holes -- many unreported trade figures -- which skew the results.  As such, we disregard these two periods from the qualitative analysis.
\item Other than the periods around the two world wars, the time period breaks naturally into two periods, roughly 1870--1975 and 1976--2006.  The transition is marked by a sharp increase in robustness in the 1970s.
\item Over these periods, robustness follows a generally decreasing trend.  The decrease in the first period (1870--1938) is statistically significant ($R^2 = 0.26, p=0.00001$ for a linear fit), the decrease in the second (1947--1974) is not ($R^2=0.07, p=0.17$), while the decrease in the third period (1975--2006) is ($R^2=0.33, p=0.0006$).
\end{enumerate}

The transition coincides with a commonly understood move away from the
\linebreak
anti-globalization policies initiated after the Great Depression \cite{dumlevcrisis}.  Our results support the claim that increased globalization corresponds to increased robustness.   But what are possible explanations for the downward trend after the transition?  An  analogous extinction analysis performed on food webs \cite{dunne2002} finds a significant positive correlation between robustness of their networks and {\em connectance}.  Connectance, defined as the proportion of existing connections in the network divided by the maximum number of possible connections (${n\choose 2}$ for a network with $n$ nodes), is a measure of the density. In the context of the WTW, if we define the relation of connectivity between nations simply to be the existence of a nonzero trade relationship, then for the full period, we find a significant {\em negative} correlation between robustness and connectance ($R^2 = 0.5, p=1.2 \times 10^{-21}$)\footnote{This $p$ value is smaller than the precision tolerance for MATLAB.}.  Moreover, we find a number of different network statistics which also have significant negative correlation with robustness in the period after transition of which we highlight one, maximum trade deficit ($R^2=0.39, p=0.0001$).

In food webs, the hypothesis is that increased connectance enhances robustness by providing, on average, more feeding opportunities for species in the face of the removal of other species \cite{d2002}.  We posit that in trade networks with our more textured (as compared with the simple ``some or nothing dynamics" of most food web simulations) dynamics  we see the opposite effect.  Extinctions create a ripple effect in the network, decreasing the income of countries who have lost exports due to an extinction as well as decreasing the availability of goods.  This ripple effect can propagate more easily and quickly through a more highly connected network.  The link we find between high trade deficits and robustness can also be understood as an aspect of this same reasoning --- imbalances between imports and exports again propagate (and potentially magnify) via our dynamics.  The impact of deleted nodes with negative trade balances (i.e., deficits) propagates primarily in the form of decreased demand which, in turn, lowers the aggregate income.

We interpret these results in terms of the trend of increased globalization since the 1970s.  Our results provide evidence for the hypothesis that globalization brings with it an number of effects, some which boost robustness of the network while other that have a negative effect.  The results point to increased connectance and the existence and rise of trade deficits as two factors that have negative effects but that coincide with the policy movement towards globalization.  These factors, in aggregate, have the effect of dramatically increasing the robustness at the outset of the policy shift with a slow decline afterwards, as the statistics associated with the negative effects grow.

For our second analysis, we consider the impact of a different sort of perturbation to the network and investigate the effect, country by country, of modulating the import/export profile of a given country.  The precise details are given below, but the basic idea is to change the import and export totals by a fixed (but different) percentage  and then to use the similar iterative dynamics as in the maximal extinction analysis to measure the impact (note that node deletion is at an extreme point in this kind of perturbation, decreasing all connections by $100\%$).  As an example, we take the 2006  WTW and  consider for each country the effect of decreasing exports by $5\%$ (for each export link) and imports by $30\%$ (for each import link).  This is meant to model a shock to a country's economy which depresses exports, such as occurred in Thailand's economic crash after the currency crises in the late 1990s where imports and exports dropped by roughly $30\%$ and $5\%$ respectively.  

Country by country, we consider the
aggregate impact on the world income (the aggregate income generated by selling exports).  In this analysis we find several things.  First, we recover,  perhaps not so surprisingly, that the large players in international trade (e.g., United States,  Germany, China, and the United Kingdom) have, in general, the largest impact, exhibiting the largest {\em power percentage} (defined as the percentage of nations whose incomes suffer at least a drop of $1\%$ as a result of the perturbation of the given nation). % Indeed, this perturbation of China creates substantial global impact according to this model, causing at least a $1\%$ drop in income for $94\%$ of countries.
However, when we consider the {\em vulnerability percentages} (defined as the percentage of nations whose perturbation {\em results} in the drop of at least $1\%$ of a given nation's  income) we see that not only are many of the weakest and/or most isolated economies  (e.g., small island nations such as Vanuatu, Papua New Guinea, the Solomon Islands, as as well as a number of South East Asian countries) most vulnerable to such perturbations , but also South American nations and larger economies such as Russia, Japan and Australia show significant vulnerability.  To examine this longitudinally, we compute the {\em maximum vulnerability percentage} --- the maximum over all countries of the percentage of countries whose perturbation results in a greater than $1\%$ decrease in income of a specific country.  This statistic exhibits a transition roughly at 1960 and in the two periods identified in our MEA analysis (again omitting the periods around the World Wars) the maximum vulnerability has a significant correlation with connectance (1870--1959, $ R^2=0.56, p=1.9 \time 10^{-14}$; 1960-2006, $R^2=0.56, p=1.6 \time 10^{-9}$).  We interpret this connection as positive relationship between connectance and robustness --- as connectance grows, the maximum vulnerability decreases.

For our third analysis, we consider the removal of bilateral trade links from the trade network and investigate their impact.  This type of experiment is meant to model situations where a trade relationship ceases -- for example, when there is a war between the two countries.  As examples, we closely analyze the impact of link deletions for a single years, 2006 and 1965, as well as more coarsely analyze the impact of link deletion for our entire range of data, 1870--2006.  For 2006 and 1965 we measure the strength of the link as the percentage of the world income lost consequent to the deletion.  We see that, in both cases, the strongest trade link is between the United States and Canada, whose deletion creates a $4.18\%$ drop in 2006 and a $3.57\%$ drop in 1965.  These two snapshots allow us to see some detailed movement over time.  In both cases, links between large economic players are the most significant.  However, we see substantial differences in these two years.  First, in 2006, a number of Asian countries such as China, South Korea, Taiwan and Japan figure prominently in the list of links with highest impact.  In 1965, however, they are largely absent (except Taiwan).  This is, of course, reflective of the growth of economic power of these countries over that period.  We also see change in the power of specific links.  For example, the extinction of the 1965 link between Mexico and the United States creates a $0.52\%$ drop in income, but in 2006 it is the second most powerful link, creating a $2.87\%$ drop.  This reflects both the growing economic power of Mexico as well as the growing intertwining of the two economies.

For a longitudinal analysis, we measure the strength of the most powerful link in each year in the same terms as well as the same statistic normalized by total link weight.  We see a trend that contrasts with the trend shown in the MEA.  Specifically, we see a general downward trend in the normalized statistic, which we interpret as a kind of {\em increasing} stability of the network in the face of this type of extinction.  However, the unweighted statistic presents more complicated behavior -- first declining with connectance and then growing as connectance continues to increase.  Moreover, we see a transition period after World War II which echoes the similar transition found by the MEA in the mid-seventies.

These give a sense of the continuum of kinds of simulations that can be performed to begin to understand the stability of the WTW in particular and perhaps, economic networks in general. Taken together, the studies that we present here provide us with a more complete picture of the stability of the WTW as a consequence of their network structure and evolution dynamics.  This combination of the topology of the network, as measured by the connectance, and its dynamics create networks which have the dual ``robust yet fragile" character.  More precisely, the MEA, which is a kind of targeted attack analysis, shows a declining robustness over time which is correlated with connectance.  The ``30/5" perturbation analysis -- again a milder form of targeted attack -- shows growing robustness over time with a sharp jump and then continued rise post World War II.  The link removal analysis shows that these networks become more and more robust to random attack over time as indicated by the decreasing influence on the world income.  Again, this measure is correlated with connectance.  However, there are still edges which, if targeted, have can have significant consequences.

Together, these studies shed light on the consequences of globalization and the liberalization of trade.  We see globalization reflected in increasing connectance -- an increasing average number of trading partners per country.  The increased connectance has two opposing consequences.  First, it correlates with decreasing robustness when using the MEA.  Second, it decreases the maximum vulnerability.  And third, on average it decreases the power of individual edges, which increases the overall robustness of the system related to edge deletions.  These results are all evidence of the claim that these trade networks are ``robust yet fragile" -- they are vulnerable to targeted attack but stable under random failure.

\section{Methods}
As our basic data source, we use import/export tables available from \cite{corrofwar} which detail the trade relationships between countries from 1870--2006 (see \cite{corrwarmethod} for details of how the data was compiled).  The resolution of the data generally gets better over time and it is worth  noting that for the periods during World War I and II  there is (not surprisingly) a good deal of missing data and we take that into account in our analysis.  These tables detail the amount of goods (in US dollars) imported or exported from one country to another and are presented as matrices which we denote as $M$ for the import matrix and $N$ for the export matrix.  Thus, $M_{ij}$ is the dollar amount of imports into country $i$ from country $j$, while $N_{ij}$ is the dollar amount of exports from country $i$ to country $j$.  While it should be the case that $M^T = N$, due to the variation in  reporting practices this is not so.\footnote{Throughout, we use the superscript $T$ to denote matrix transpose.}  For simplicity, we analyze only the import matrix $M$ and in this way define the weighted directed WTW.  %Our basic analytic object is the World Trade Web which we build from the export matrix as a directed graph.  We place an edge between node $i$ and node $j$ if $E_{i,j} \neq 0 $.

%This data is collected and aggregated from a number of sources and suffers  in some years, mainly the early ones, from substantial missing data, and in general, the data is more complete over time. More

 % and obvious - many of the world's largest economies at the time make no report of trade between themselves and their former (and later) large trading partners.
%This is perhaps unsurprising as we suspect that many of these countries had higher priorities than import/export reporting in these periods.  But, as the resolution of the data is particularly low during these periods, we cannot infer much from our analyses during these periods.  As such, we exclude them from the interpretive analysis.  However, we can still consider them as abstract networks from the point of view of attempting to find relations between network statistics and robustness.

 %More specifically, let   $S = \{i_1,\dots, i_k\}$ denote a subset of the nodes in $M$ (as usual, we identify $M$ with the set of underlying nodes). Suppose we delete $S$. Then node $j$ has
%$$exc_S(j)=\sum_{\ell\in S}M_{j\ell}$$
% excess dollars of exports (relative to $S$) to redistribute.  Similarly, if node $j$  imports from one or more of the nodes in $S$, then with the deletion of $S$ it has a new unmet demand arising from the deletion defined as
% $$dem_S(j)=\sum_{\ell\in S} M_{\ell j}.$$

 \subsection{Income model}\label{Methods}
 Any sort of extinction analysis in the WTW requires the definition of the network consequences of node removal. Our working assumptions are as follows:
\begin{itemize}
\item[(1)] Node deletion creates an ``excess" of dollars (supply) in the global economy given by the goods that a deleted country would have purchased from other countries.
\item[(2)] Node deletion creates an unmet demand in the form of goods that other countries previously purchased from a deleted country.
\end{itemize}

To understand the effect of the removal of one or more nodes, we need to instantiate dynamics on the network to model how the trade network rebalances trade after such a shock, while taking into account the above assumptions.  To do so, we introduce an {\em income model} on the network.  The basic economic assumption is that as a country's income from selling its exports increases, its demand for products (in the form of imports) increases as well. Thus,   the deletion of a country has the effect of both causing a decrease in the demand and supply for its direct trading partners,
but also an increase in demand for other countries via the first order unmet demand of its trading partners.

As a first step toward concretely modeling this dynamic, we define a country's {\em propensity to spend} based on the in-degree and the out-degree of the original import matrix, whose definition we recall here.

 \begin{Def}
Let $M$ be an import matrix.  Then, the {\em in-degree} for node $i$ is \[D_M(i)=\sum_{j}M_{ji}\]
and the {\em out-degree} for node $i$ is
\[O_M(i)=\sum_j M_{ij}.\]
\end{Def}
In short, the in-degree $D_M(i)$ is simply the total dollar value of goods imported by other countries from country $i$.  The out-degree $O_M(i)$  is the total dollar value of goods imported by country $i$ from all other countries.  For our purposes, the in-degree gives a measure of the income of country $i$ while the out-degree gives a measure of its expenditures.

We define {\em the propensity to spend} for country $i$ as
\begin{equation}\label{model1}
\alpha_i=
\begin{cases}
\frac{O_M(i)}{D_M(i)} \;\;\text{ if $D_M(i)\ge O_M(i)$}\\
1 \;\; \qquad \; \qquad \;     \text{otherwise}
\end{cases}
\end{equation}
In the case when a country spends more than it earns ($O_M(i)>  D_M(i)$) we assume that the internal economy of the country is producing excess dollars (by some means) that are used to fund additional import spending.  We record this as
\begin{equation} \label{model2}
\beta_i=
\begin{cases}
O_M(i)- D_M(i) \;\;\text{ if $O_M(i)>  D_M(i)$}\\
0 \;\; \qquad \; \qquad \;     \text{otherwise.}
\end{cases}
\end{equation}

Next, we define a stochastic matrix $m$ associated to the import matrix $M$ via
\begin{equation}\label{model3}
m_{ij}= M_{ij}/O_M(i).
  \end{equation}
  We call $m$ the {\em propensity to import} matrix.  In mathematical terms, it is simply the Markov chain associated to $M$.
  Note that with these definitions,
\begin{equation}\label{model4}
\begin{split}
O_M(i)&= \alpha_i D_M(i) + \beta_i\\
M &=  diag(O_M)m\\
\end{split}
\end{equation}
where for a vector $v$,  $diag(v)$  is the diagonal matrix with the entries of $v$ along the diagonal.

We now model the iterative buying and selling between trade partners as a two-step updating rule.  Let the import matrix at time $t$ be denoted as the matrix $M_t$.  Then, the vector of incomes $I_t$ is given by
\begin{equation}\label{model5}
I_t = diag(\alpha) M_t^T \bf{1}+\beta
\end{equation}
where $\alpha = (\alpha_1,\dots,\alpha_n)^T$, $\beta=(\beta_1,\dots,\beta_n)^T$ and $\bf{1}$ $=(1,\dots, 1)^T$.  To  iterate  we calculate the next import matrix from the income:
\begin{equation}\label{model6}
M_{t+1}=diag(I_t)m.
\end{equation}
We can then repeat, forming $I_{t+1},M_{t+2},\dots$.  Note that from \eqref{model4}, we see that for import matrix  $M$
with associated quantities $\alpha, \beta$ and $m$ (derived from $M$ as above), then $M$ (and the resulting income associated with $M$) is a fixed point of this iteration.  Moreover, if we fix $m,\alpha,$ and $\beta$, we can solve the system for fixed points, producing the equilibrium income as
\[ I = -(diag(\alpha)m^T - \mathbb{I})^{-1}\beta\]
where $\mathbb{I}$ is the identity matrix.\footnote{The invertibility of the matrix is equivalent to the fact that $diag(\alpha)m^T$ does not have a fixed vector.  This is true in most cases because the maximum eigenvalue of $m^T$ is $1$ and each entry in  $\alpha$ is at most $1$.  The only case where this isn't true is if $\alpha=\bf{1}$.  In that case, $m^T-\mathbb{I}$ is not invertible, but $(m^T-\mathbb{I})I=-\beta$ can still be solved so long as $\beta$ is not a multiple of the constant vector.}  Note that in a limiting case where $O_M=D_M, \alpha=(1,\dots,1)^T,\beta=(0,\dots,0)^T$ the equilibrium income is simply the fixed vector of $m^T$.  It is in this sense that we see the income model as a perturbation of a Markov process.

From this we can describe how the income model responds to shocks to the system (defined as any perturbation of the network) via the following algorithm:
\begin{enumerate}
\item Let $M_0$ be the initial import matrix and calculate $\alpha, \beta$ and $m$ from \eqref{model1},\eqref{model2}, and \eqref{model3}.  Let $I_0$ be the unshocked income vector given by \eqref{model5} using $M_0$.
\item Let $M_1$ be a modified version of $M_0$ encoding the desired shock.  Adjust $\alpha,\beta$ and $m$ to reflect the modification (this will be detailed in each specific case below).  Let $t=1$.
\item Calculate $I_t$ via \eqref{model5}.
\item Calculate $M_{t+1}$ via \eqref{model6}.
\item Increment $t$ and repeat the last two steps.
\end{enumerate}

Some comments are in order.  First, we emphasize that $\alpha, \beta$ and $m$ are fixed for the iteration steps of the simulation.  We envision the simulation as an approximation of a short time rebalancing.  Thus, we do not allow the fundamental constants --- the profile of countries that a country trades with and in what proportion, the propensity to spend and the amount of internal income --- to change.  Second, we have an explicit assumption of {\em substitutability}.  We recognize the shortcomings: imports and exports of wildly different items are lumped into the aggregate statistics. Analogous assumptions are made for extinction models for food webs -- prey are assumed to be interchangeable (and available) for a given predator species.  In our setting, this assumption provides the easiest path of rebalancing the system after a shock --- thus providing the network with the best possible chance of recovery. In addition, one of our assumptions in the model dynamics helps mitigate the assumption of substitutability:  as we do not allow countries to form new trading partners during the simulation, countries cannot secure available goods from other countries but only more (or less) goods from their original partners.  As the majority of countries have multifaceted trading relationships --- buying one good from many different countries --- this is a proxy for the availability of substitutable goods from existing trading partners.  This is particularly true of the larger economies in each network which, as we see in the simulation, have the most impact on its stability.

\subsection{Shocks via node removal}

In our investigations, we first consider a shock created by disconnecting a node from the network.  This is ``node removal" or ``node deletion." Using the dynamics framework, the disconnection of node $i$ is effected by constructing $M^i_1$ from $M_0 = M$  by setting  $M_{i r}$ and $M_{r i}$ equal to zero for all $r$.
%This preserves the number of nodes in the network representation of the matrix (and allows for the matrix multiplication steps in the algorithm above), but changes the connection structure.

A {\em maximal extinction analysis} is performed by carrying out the following algorithm:
\begin{enumerate}
\item Fix the initial import matrix $M$ and its associated income matrix $I$.  For the first iteration, let $M_0=M$, $I_0=I$.
\item For each $i$, delete node $i$ from $M_0$ to form $M_1^i$. Adjust $m$ by zeroing out the $i^\text{th}$ row and column and renormalizing.  Adjust $\alpha$ by setting $\alpha(i)=0$ and $\beta$ by setting $\beta(i)=0$.
\item Calculate $\{M_0,M_1^i,\dots,M_5^i\}, \{I_0,I_1^i,\dots,I_5^i\}$ via the income model.\footnote{We also repeated this procedure with 10 and 50 iterations, instead of 5.  The results were qualitatively the same.}
\item Calculate  {\em total power of node $i$} defined as
\[Pow(i) = 1-\frac{\sum_{j} I_5^i(j)}{\sum_{j} I_0(j)}.\]
It represents the percent of total income left after node deletion and rebalancing.  Note:  sometimes this number is negative, signifying that the node deletion results in a net increase in income.
\item Find the node $j$ so that the total power is maximized:
\[j = \arg \max_{k} Pow(k)\]
Add this node index to the list, $\mathfrak{D}$, of deleted nodes.
\item  Letting $M_0=M_5^j$ and $I_0=I_5^j$.  Repeat steps $2-5$ this until the total income falls below $50\%$ of the total income of the original import matrix, i.e.
    \[ \frac{\sum_{k} I_5^j(k)}{\sum_k I(k)}<0.5\]
\end{enumerate}

From this algorithm, we measure {\em robustness} as \[R= \frac{ \sum_{j \in \mathfrak{D}} I(j)}{\sum_k I(k)},\]
the percentage of the initial income removed to reach, according to these dynamics,  the $50\%$ threshold.

From its definition, $0 < R \le 0.5$.  Higher (resp. lower) values of $R$ correspond to higher (resp. lower) robustness of the network to the maximal extinction process. This mirrors the food web framework \cite{d2002} in which robustness is defined as the minimal fraction of species that need to be deleted (where the dynamics proceeds by removing a node when all of its food sources have disappeared) in order to destroy at least $50\%$ of the species in the ecosystem.

The removal of an entire node is meant to create a ``worst-case scenario"  -- the removal of an entire country from the import-export economy.  While this may seem extreme, it does provide a direct way of measuring the impact and power of a particular country over the entire system.  Similarly coarse assumptions are used in robustness analysis for food webs -- for example, prey are assumed available in sufficient quantity to sustain predator populations until extinction. As the model is inherently linear, we would find similar results if we instead removed only a portion of a given country's import/export profile. In fact, in the subsequent experiments we use these sorts of modifications  to arrive at different measures of economic power.

 \subsection{Shocks via node perturbation}
In this section, we present a more textured form of shock based on edge perturbation,  rather than the deletion of an entire set of incoming and outgoing edges at a given node.  This represents a  model for intrinsic or extrinsic shocks to economies of individual countries which (temporarily) affect their import/export profile.  Good examples   can be found in the changes that occurred in some of the southeast Asian economies during the currency crises of the 1990s.  For example, Thailand enjoyed rapid growth in the late 1980s through much of the 1990s -- making it a so-called ``tiger" economy -- but rapidly declined due to a crisis in the value of the bhat in 1997--1998. This brought massive unemployment and economic hardships to the nation and with that,  significant drops in the value of exports and  in imports  (by $30\%$ and   $5\%$ respectively).

 To model such a shock, we perturb the WTW by a fixed manipulation of a specific node's import and export values. For example, we can create a two-parameter perturbation of imports and exports.  If we fix scale factors $a,b \in [0,1]$ and a node $i$, we define
 \begin{equation*}
 (M_1)_{jk} =
 \begin{cases}
 (M_0)_{jk} \;\;\text{if $j,k \neq i$}\\
 a(M_0)_{ik} \; \; \text{ if $j=i$}\\
 b(M_0)_{ji} \; \; \text{ if $k=i$}.\\
 \end{cases}
 \end{equation*}
 Note that other more complicated perturbations to the import/export profiles could be adopted (e.g., independent modulation of the edges).

 Using this $M_1$, we must then adjust $m,\alpha,$ and $\beta$ accordingly.  For $m$, we scale the $i^\text{th}$ row by $a$ and then $i^\text{th}$ column by $b$ and renormalize.  The propensity to spend vector $\alpha$ is changed only in the $i^\text{th}$ entry which is scaled by $\frac{b}{a}$.  We leave $\beta$ unchanged.  The choice to leave $\beta$ unchanged is reflective of our desire to produce a model which encodes short term response to shocks.

 Next, we follow the algorithm in Section \ref{Methods} to compute the impact.  As we noted above, if $a=b$, the results will be proportional to the full extinction of a node due to the linearity of the dynamics. 
 %When $a\neq b$, this model provides a method by which to evaluate test hypotheses regarding different perturbations.

\subsection{Shocks via link removal}
We can also model shocks created by the termination of a trade relationship.  In this case, we simply remove a trade link by deleting all edges from the import/export matrix associated with a bilateral trade relationship.  This models a situation where a trade relationship is completely dissolved. Natural examples include wars, trade embargoes, etc.
If we denote the two countries involved by $i$ and $j$, then the shock is encoded as
\begin{equation*}
 (M_1)_{kl} =
 \begin{cases}
 (M_0)_{kl} \;\;\text{if $(k,l) \neq (i,j),(j,i)$}\\
 0 \; \; \qquad \;  \text{if $(k,l) = (i,j),(j,i)$}.\\
 \end{cases}
 \end{equation*}
 With this $M_1$, we then adjust $m$ by deleting the same entries and renormalizing the matrix.  The values of $\alpha$ and $\beta$ are unchanged.  Then, following the algorithm in Section \ref{Methods}, we compute the impact.
\section{Results}
\begin{figure}
\begin{center}
\includegraphics[scale=0.75]{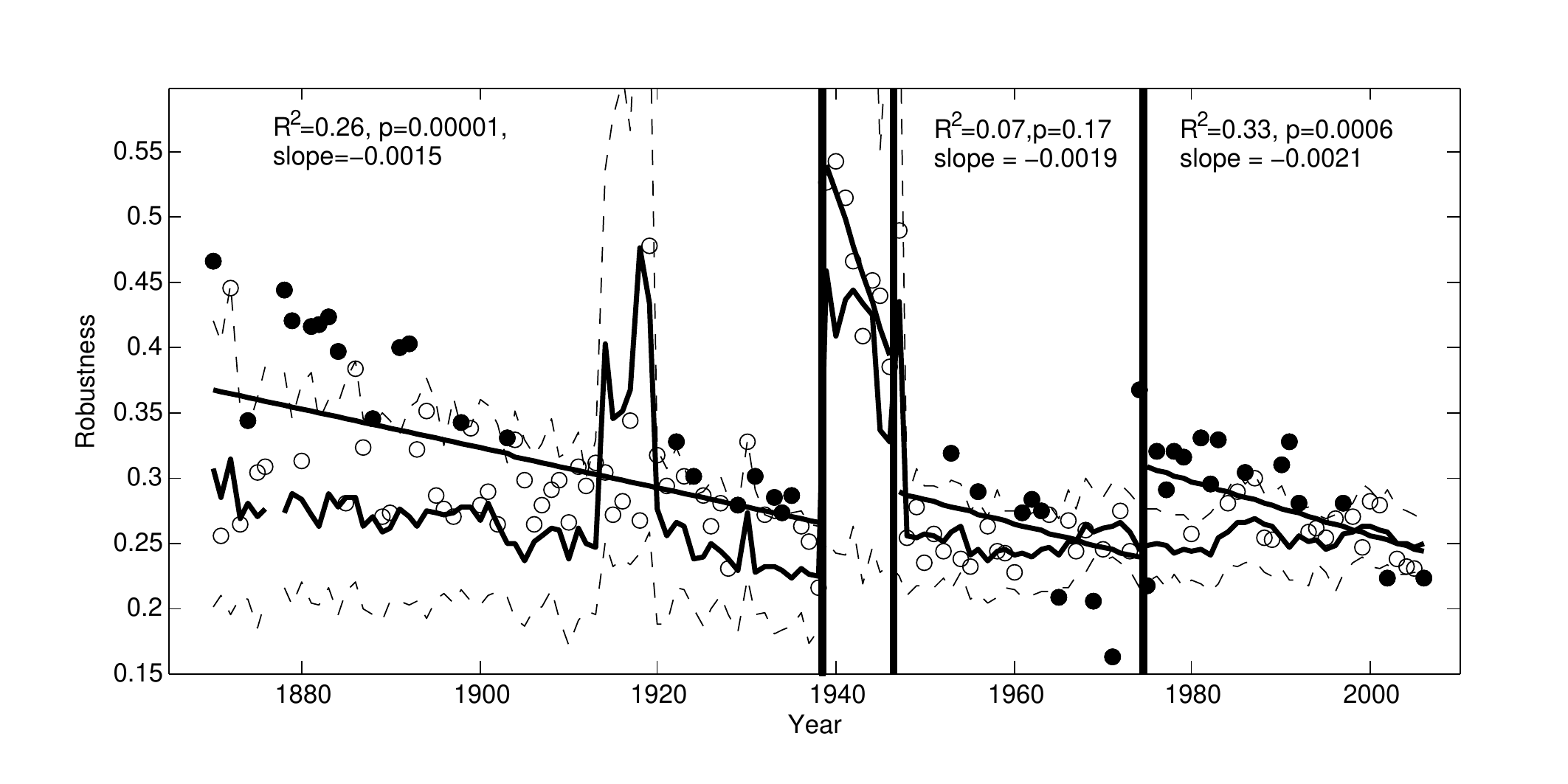}
\caption{Robustness scores for the WTW over time.  Solid lines denote mean robustness of $50$ trials of null models, dotted lines are the $95\%$ and $5\%$ cutoffs.  Filled circles show years with robustness scores that fall outside the $5-95\%$ interval.}\label{fig1}.
\end{center}
\end{figure}
Figure \ref{fig1} shows the results of the robustness computation under the maximal extinction analysis over time.  Outside of the years around World War I and II, we see that there are two regimes roughly  split at 1975.  Before the mid-seventies, the robustness is low and decreasing over time.  In the seventies, we see a rapid increase in robustness, and then the resumption of the decreasing trend.  In the period around World War II, we see a sharp increase in robustness.  As stated above, this is due to the sparsity of the reported data rather than related to an actual economic change.  We note that around World War I, we have similar sparsity.  But, in that case, while we see an uptick in robustness, the transition is not nearly so sharp.

To better understand the significance of these results, we compare the robustness measurements to an appropriate null model.  For the null model, we use a randomization of the import matrix (i.e., permuting the endpoints of the outgoing edges at each node) and repeat our maximal extinction analysis on the result.  Repeating this multiple times provides a family of null models for a given import matrix and a corresponding distribution of robustness scores.  Figure \ref{fig1} shows the robustness scores from 1870--2006 plotted with the mean, minimum, and maximum of robustness scores of $50$ runs of each null model.  The robustness scores are coded by shape to indicate their significance.  The scores shown as filled circles are either larger than $95\%$ or smaller than $5\%$ of the corresponding scores for the $50$ null models.  Thus, according to this threshold, these robustness scores are significantly different than randomized null models with the same number of nodes and total out-degree.

We can interpret this as follows.  In the periods from 1870--1913 and 1920--1939, some aspect of the structure of the networks creates higher robustness than expected by chance in a number of years. Overall, we see a slow decline of robustness over time.  We then see a similar picture of decline between 1949 and 1975, with a transition in 1975--1976.   Shortly after the transition, the robustness increases to a level significantly above that of comparable null models but then again begins a decline.  In the first periods, the downward trend is statistically significant ($R^2=0.26, p=10^{-5}$) while in the period from 1949--1975, it is not significant ($R^2=0.07,p=0.17$).  In the last period, 1976--2006, the trend is again significant ($R^2=0.33, p=6 \times 10^{-4}$).  To attempt to understand the decline over time, we consider statistics with potential predictive power.  Following the food web literature, we consider the relation to {\em connectance}, which is simply the ratio of edges in the network to possible edges in the network \cite{d2002}. In our case, we count the ratio of nonzero trading relations to the total\footnote{In truth, we should not allow for the possibility of a node to be connected to itself, but this overcounting has negligible effect on our result.} possible
\[C= \frac{\text{\# of nonzero edges}}{(\text{\# of nodes})^2}.\]
We also consider the relation to the {\em maximum trade deficit} defined as the maximum of $D_M-O_M$ for a given import matrix $M$.

\begin{figure}
\begin{center}
\includegraphics[scale=0.75]{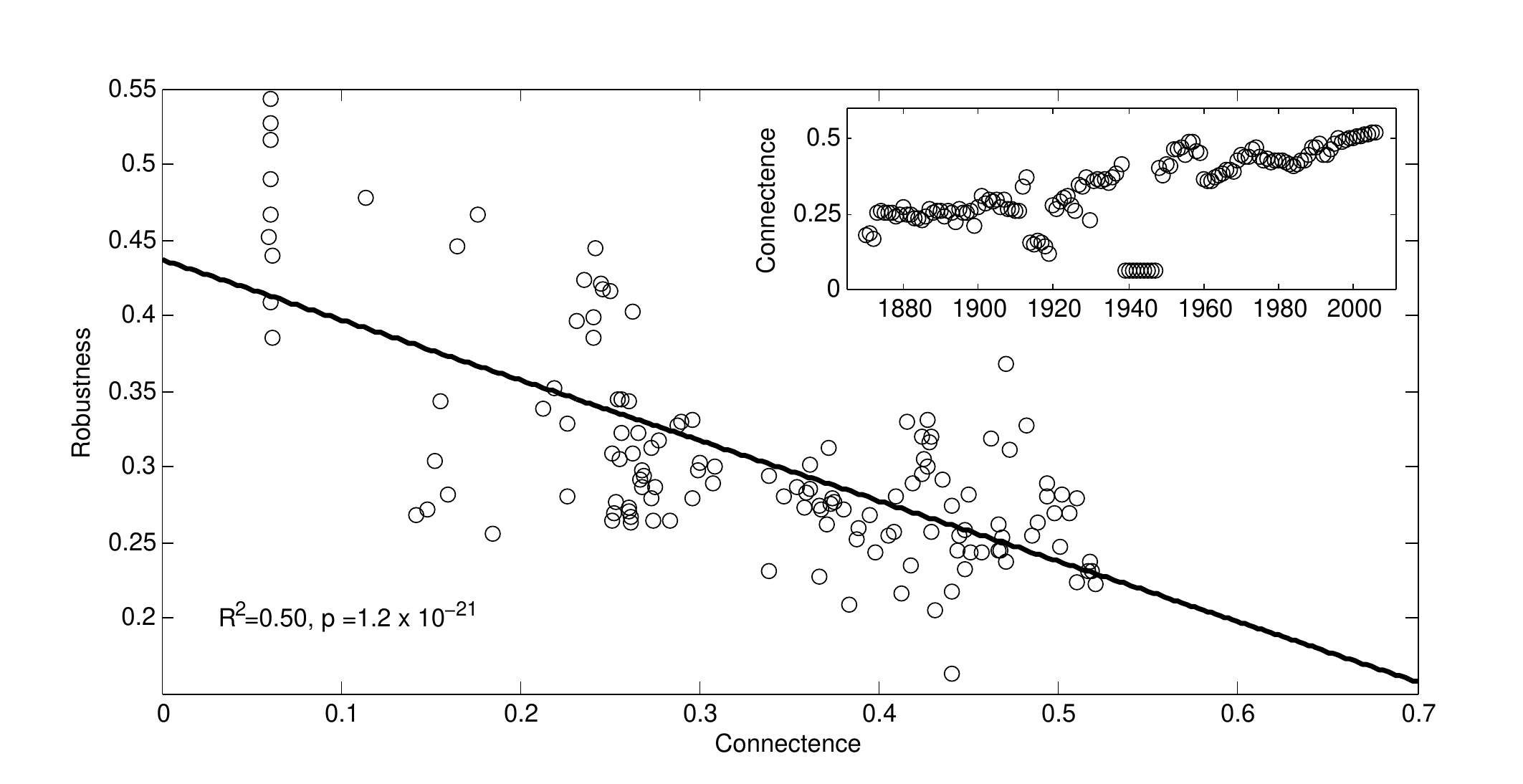}
\caption{Plots of robustness vs. connectance  over the period 1870--2006.  The inset graph show the time series for connectance for 1870--2006.}\label{fig2}
\end{center}
\end{figure}
In Figure \ref{fig2}, we plot the robustness against connectance. The inset graph shows the plot of the connectance over the entire time period, 1870--2006.  We also plot the best linear fit of the data and provide the $R^2$ and $p$-values for the regression.  The fit is statistically significant with $R^2=0.5$ and $p=1.2 \times 10^{-21}$.  In this computation we include the periods around both World Wars.  We do this as we are merely attempting to link the summary statistics of robustness and connectance and are not (yet) interpreting the results in terms of actual trade relationships.  So, as the trade webs in these periods are simply incomplete trade webs, they are still appropriate for inclusion in this analysis.   We again note that the sharp drop in connectance around the World Wars is likely due to the sparsity of the data and the resulting incomplete trade web.  We also note the smaller drop in connectance in the early 1960s.  This corresponds to a sudden increase in the number of countries in the trade webs which, in turn, is a consequence of the wave of former colonial countries gaining their independence.  Generally, as these countries gained independence they entered the world trade network slowly -- first trading with geographic neighbors and their former colonizing country.

It is interesting to note that while we observe a negative correlation between robustness and connectance in the WTW, a positive correlation is observed in food webs. One possible explanation comes from the difference in dynamics. A maximal extinction analysis for food webs includes no population dynamics (see e.g., \cite{Allesina2009}).  After a targeted extinction, consequent extinctions in the network only occur if a species no longer has any prey (except, possibly, itself).  In our situation, the income model provides simple linear dynamics associated with the fundamental principles of supply and demand.  These dynamics are one type of analogue to population dynamics on food webs.  They allow for the shock associated with node removal to propagate through the network, much like a contagion.  As higher connectance generally implies faster propagation speeds through a network, it is not surprising that we see a significant negative correlation.

\begin{figure}
\begin{center}
\includegraphics[scale=0.75]{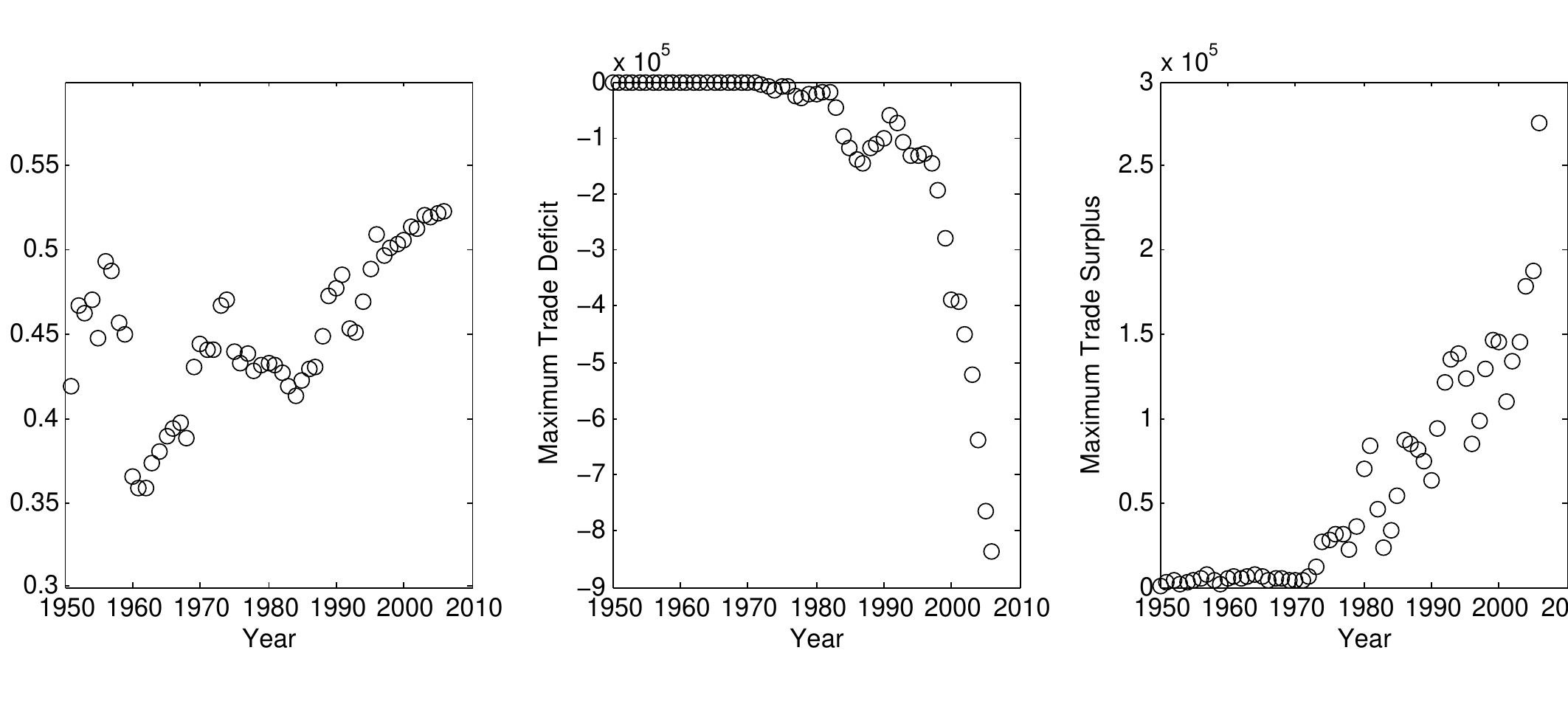}
\caption{Plots of connectance, maximum trade deficits and maximum trade surpluses from 1950--2006.}\label{fig3}
\end{center}
\end{figure}

 A similar argument applies as a possible explanation of the positive correlation between the maximum trade deficit and robustness.  The deletion of a node with a substantial deficit again propagates through the network but the deficit itself creates unused supply which is not balanced (overall), by unmet demand.  Irrespective of the network topology, this creates downward pressure incomes.  %If we were to consider maximum trade surplus instead, a similar argument indicates that, irrespective of network topology, we would expect an upward pressure on incomes.  The fact that maximum trade surplus has a weaker but significant correlation ($R^2=0.25, p=0.01$) seems to contradict the previous argument. However, we are likely seeing interconnected and interdependent measurements.
The jump in the 1970s and then decline in robustness over time tracks the increased connectance and the existence (and increasing size) of trade imbalances.  The regression analysis for robustness in terms of maximum trade deficit yields a significant result ($R^2=0.39, p=10^{-4}$).  It seems plausible that these are linked with a move towards policies of increasing globalization.  As we see in Figure \ref{fig3}, the effects of these changes on maximum trade deficits and maximum trade surpluses are relatively mild at first before rapidly growing. The connectance decreases for a time before assuming an upward trend. Thus, one conclusion we may draw is that the positive effects on robustness are eventually mitigated by the negative consequences of the other changes to the network.  This is one plausible explanation for a significant reorganization initiating a  peak of robustness following followed by a  steady decline.

As an example of the second methodology, shocks via node perturbation, we considered the 2006 import/export matrix and perturbed each node sequentially with model parameters $\alpha =0.7, \beta = 0.95$. In other words we modeled a $30\%$ drop in imports and a $5\%$ drop in exports (motivated by the drops seen in the late nineties for Thailand as a result of their currency crisis)
and analyzed the impact according to the model. To create an aggregate statistic to help visualize the results of the simulation, for each country, we counted the number of other countries whose income decreased by $1\%$ or more when we simulated the drop in imports and exports.  If we divide this count by the total number of countries, we call the result the {\em power percentage}.  We also calculate the {\em vulnerability percentage}, which is the number of countries who perturbation create a $1\%$ or more decrease in income of a given country divided by the number of countries.

\begin{figure}
\subfigure[Power Percentage]{
\includegraphics[scale=0.35]{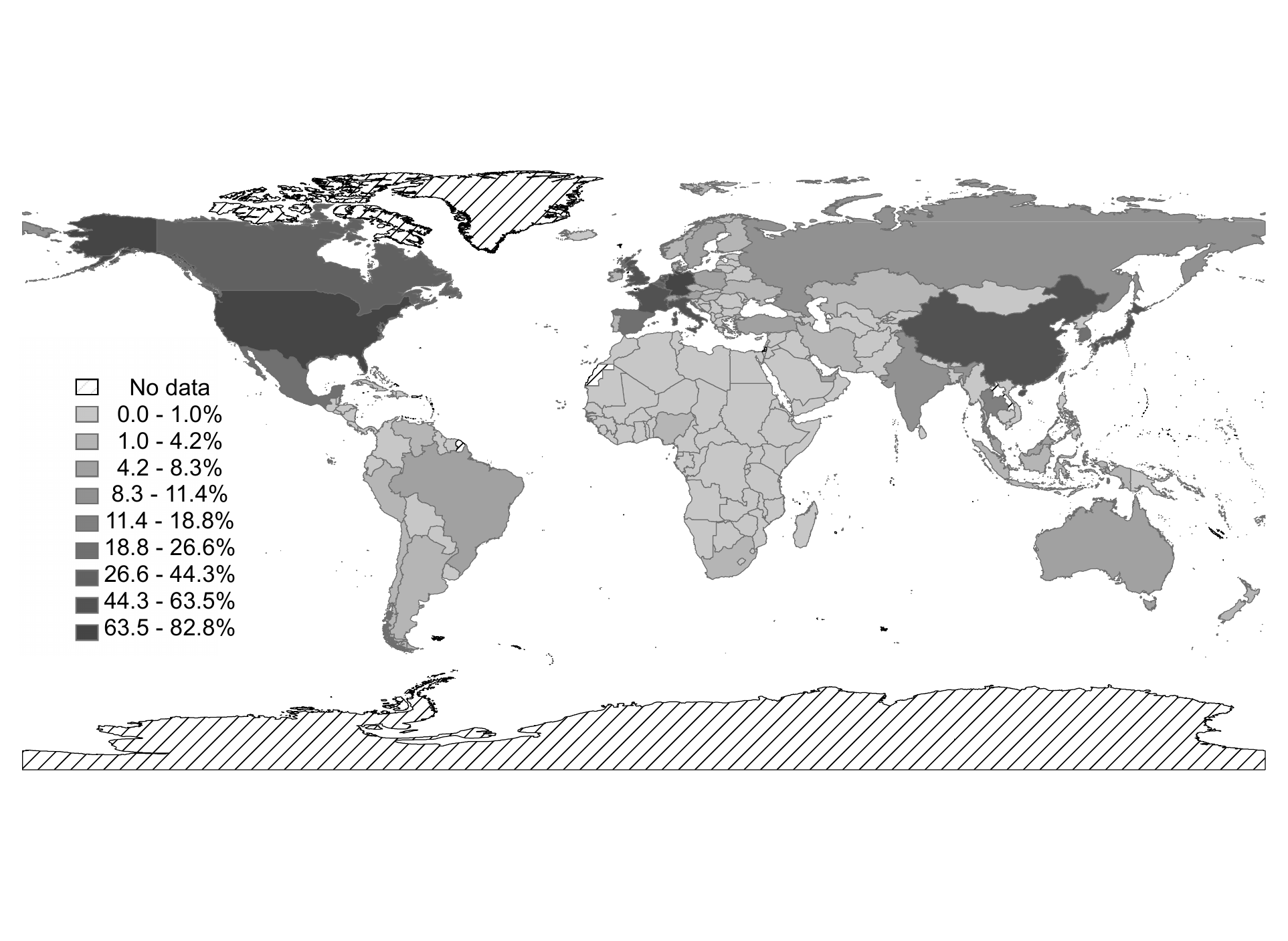}
}
\subfigure[Vulnerability Percentage]{
\includegraphics[scale=0.35]{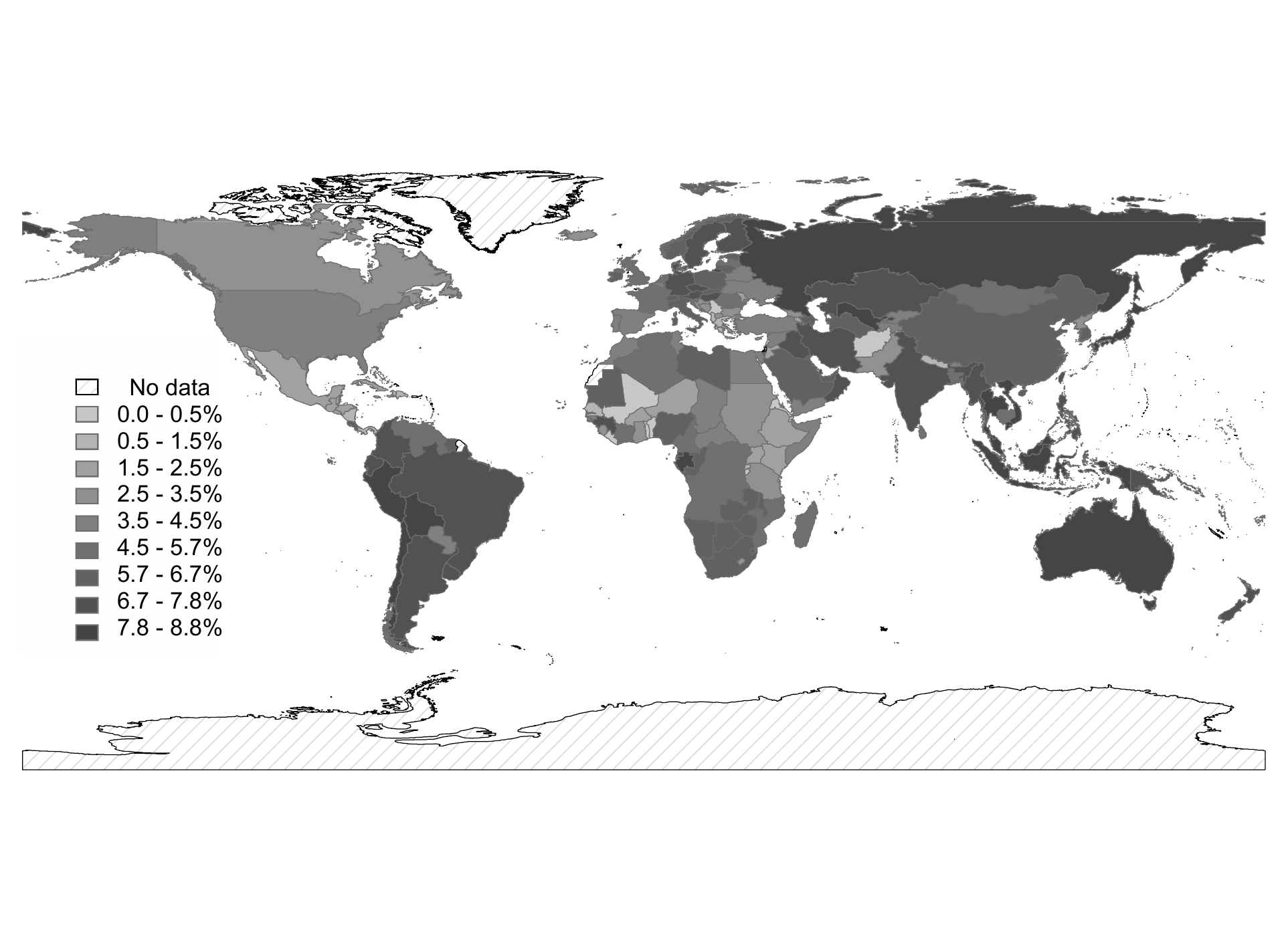}
}
\caption{Two world maps with coloring indicating the power percentage and the vulnerability percentage for the WTW in 2006.  Hatched countries are ones where no data was available for the computation.}\label{worldmap}
\end{figure}

An initial review of the results show us something relatively unsurprising, that the largest players in the world economy - the United States, Germany, China, the United Kingdom, the Netherlands, France, Japan, Italy, Canada, and Belgium - have the most impact .  Somewhat more surprising is the list of most vulnerable countries, which include a number of southeast Asian countries, South American Countries, Japan, Australia, and Russia (see Table~\ref{tbl:pp}).  Figure \ref{worldmap} illustrates this using world maps.  The left map is colored by the power percentage while the map of the righthand side is colored according to the vulnerability percentage.

\begin{center}
\begin{table}
\begin{tabular}{|l|c||l|c|}
\hline
{\bf Country} & {\bf Power Percentage} & {\bf Country} & {\bf Vulnerability Percentage}\\
\hline
U.S.A. & 82.81 & Uzbekistan & 8.85\\
\hline
Germany & 78.65 & Thailand & 8.85\\
\hline
China & 63.54 & Indonesia & 8.85\\
\hline
U.K. & 62.50 & Peru & 8.85\\
\hline
Netherlands & 62.50 & Bolivia & 8.33 \\
\hline
France & 61.46 & Chile & 8.33\\
\hline
Japan & 53.13 & Russia & 8.33\\
\hline
Italy & 52.60 & Gabon & 8.33 \\
\hline
Canada & 44.27 & Japan & 8.33\\
\hline
Belgium & 40.10 & Vietnam & 8.33\\
\hline
\end{tabular}
\smallskip
\caption{{\small On the lefthand side we list the top ten countries in terms of power percentage for the 2006 WTW. The next column is the percentage of countries that experience at least a 1\% drop in  income due to decreasing a 5\% drop in each export link and a 30\% drop in each import link.  The righthand side of the table shows the list of the top ten most vulnerable countries for the 2006 WTW according to vulnerability percentage under the same type of perturbation.}}\label{tbl:pp}
\end{table}
\end{center}

\begin{figure}

\subfigure[Vulnerability Percentage]{
\includegraphics[scale=0.35]{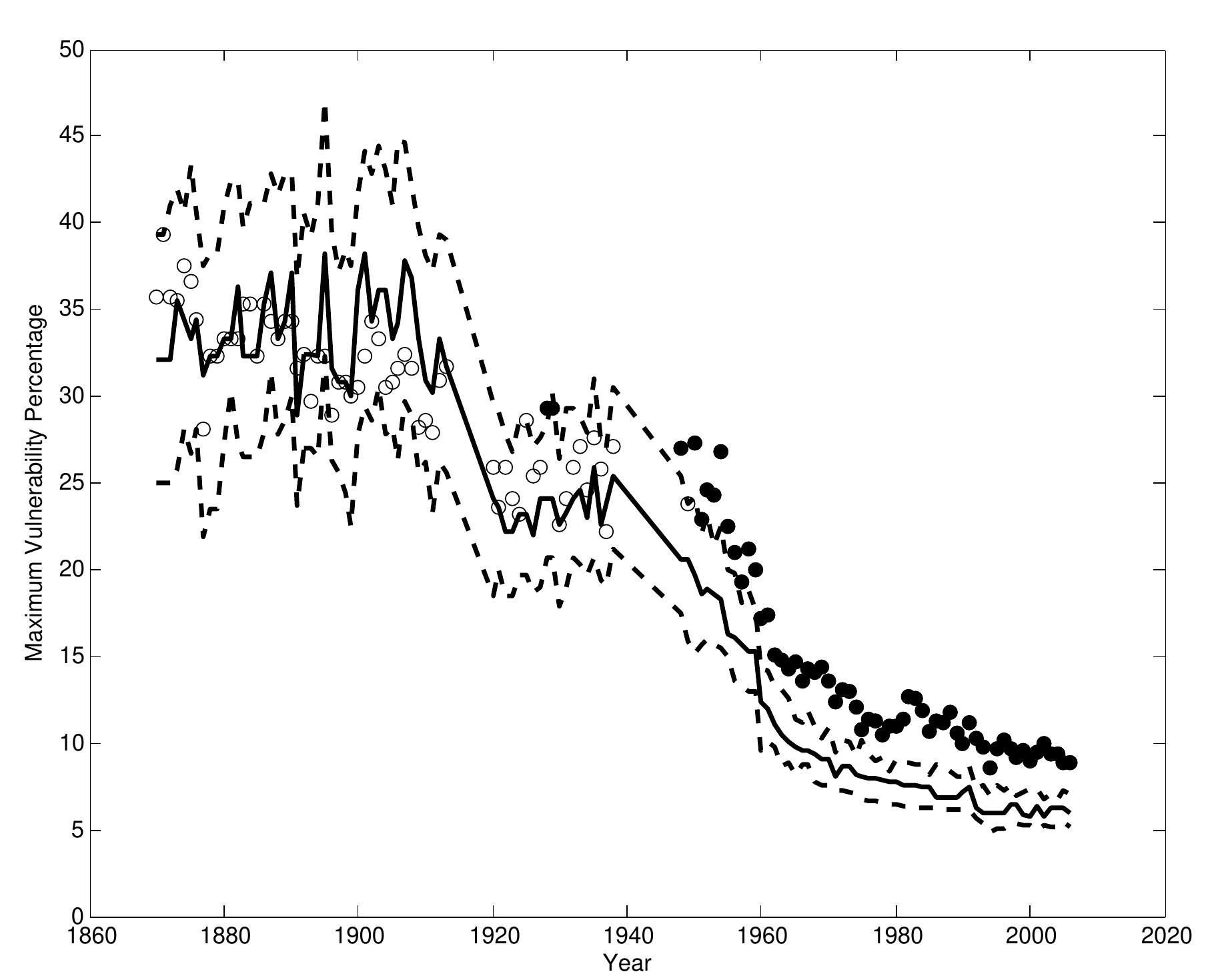}
}
\subfigure[Vulnerability vs. Connectance]{
\includegraphics[scale=0.5]{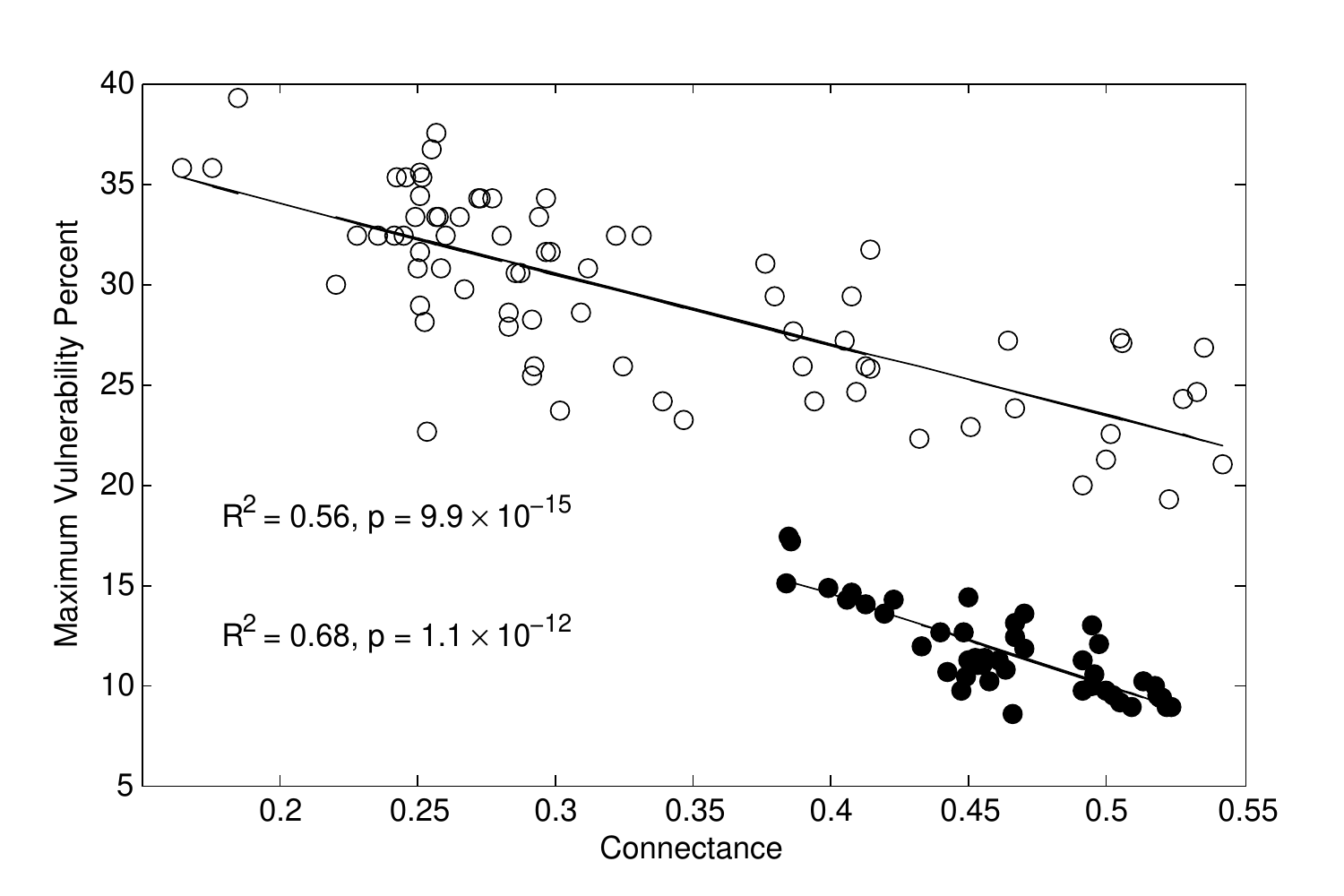}
}
\caption{(a) Vulnerability Percentage over time.  The solid line is mean percentage over 100 null model trials while the dotted lines are the $5\%$ and $95\%$ cutoffs from the null models.  The circles in black are years where the vulnerability percentage is outside the $5-95\%$ range.  (b) Vulnerability percentage plotted against connectance.}\label{percfig}
\end{figure}

%This map shows that a small number of economies hold substantial power over the world trade web.  Moreover, most countries are vulnerable to perturbations of their trading partners and some groups of countries - notably those listed above - are especially vulnerable due to their patterns of trade.
%
To get a sense of this method applied over time, we repeat the same experiment for each year in our data set and calculate the maximum power and vulnerability percentages for each year.    Figure \ref{percfig} shows these results. Note that in each case, we have omitted the scores from the periods around the two World Wars, due to data sparsity.  In computing the maximum power percentage over this time period, we find it is relatively uniform and large -- always above $75\%$ and mostly above $90\%$.  In Figure ~\ref{percfig}(a) we see that the maximum vulnerability percentage decreases over time, which we interpret as growing robustness of the system.  We also note the transition after World War II which echoes the transition found in the MEA. In investigating the results of the MEA, we looked at the relationship between connectance and robustness.  We do the same here (Figure~\ref{percfig}(b)) and find a strong correlation between connectance and the maximum vulnerability percentage.  However, there are two distinct regimes which amplify our earlier observation of a transition -- the black circles are years after 1960 while the white circles are before.  In this experiment, we interpret this as linking growing connectance with growing robustness as measured by the maximum vulnerability.   This interpretation dovetails with the analysis of the relationship of connectance and the MEA.  For the MEA, we found that growing connectance decreases robustness when measuring cascading extinctions.  On the other hand, growing connectance increases robustness from the point of view of the maximum vulnerability.  This is an aspect of what are often called ``robust yet fragile" networks -- those which are fragile in the face of specific (targeted) attack but stable in the face of nonspecific (random) attack.

As another example of a kind of perturbation analysis we consider the importance of indivdual trade relationships in the WTW. We begin by looking at the 2006 WTW and systematically delete edges associated with individual bilateral trade relationships.  Table \ref{edgetab} shows the removals that had greater than a half-percent negative impact on aggregate world income.
\begin{table}
\begin{tabular}{|c|c|c|}
  \hline
 Country 1 & Country 2 & Percent change \\ \hline
Canada & United States of America & -4.18\\ \hline
Mexico & United States of America & -2.87\\ \hline
Germany & Netherlands & -1.03\\ \hline
Germany & France & -1.03\\ \hline
China & United States of America & -0.96\\ \hline
Japan & United States of America & -0.92\\ \hline
United Kingdom & United States of America & -0.87\\ \hline
France & Belgium & -0.82\\ \hline
Belgium & Netherlands & -0.81\\ \hline
Germany & United States of America & -0.80\\ \hline
Germany & Belgium & -0.79\\ \hline
Japan & China & -0.72\\ \hline
Italy & Germany & -0.72\\ \hline
Germany & United Kingdom & -0.72\\ \hline
Italy & France & -0.70\\ \hline
Spain & France & -0.67\\ \hline
South Korea & China & -0.62\\ \hline
France & United Kingdom & -0.61\\ \hline
France & United States of America & -0.55\\ \hline
Taiwan & China & -0.54\\ \hline
Austria & Germany & -0.53\\ \hline
South Korea & United States of America & -0.52\\

 \hline
\medskip
\end{tabular}
\caption{The results of shock via link deletion for the 2006 trade web.}\label{edgetab}
\end{table}

We again see the impacts dominated by large economic players - the United States, European countries, China, South Korea, Taiwan and Japan.  But, as with the our other analyses, a more nuanced picture appears with closer analysis.  For example, the links between the United States and Canada and between the United States and Mexico are the most powerful by this metric.  While this is a consequence of the dominance of the United States in the world economy, it also reflects the impact of the tighter integration of North American economies due to the general liberalization of trade as well as the free trade agreements put in place in the 1990s.  This is exhibited by the strength of these trade ties which, while not the largest ties in the world economy, have the most aggregate impact.  To see this more clearly, we compare the results from 2006 to those of the same type of link removal for the year 1965.  Table \ref{edgetab2} again shows the link with negative impact larger than half a percent.  While we see the same basic pattern, ties including the largest economic players have the largest impact, there are also interesting changes.  The most stark is the change in predominance of the link between Mexico and the United States.  In 2006, it is has the second largest impact of $-2.87\%$ while in 1965 is the last on this list with $-0.52\%$ impact.  One conclusion we can draw is that the move towards liberalized trade policies has two contradictory effects:  a general robustness (in the sense that impacts generally decline) accompanied or  tempered by vulnerability evidenced by the much higher impact of specific trade partnerships.
 \begin{table}
\begin{tabular}{|c|c|c|}
  \hline
Country 1 & Country 2 & Percent Change \\ \hline
Canada & United States of America & -3.57\\ \hline
Germany & France & -2.22\\ \hline
Germany & Netherlands & -1.99\\ \hline
Taiwan & United States of America & -1.86\\ \hline
Hungary & Germany & -1.48\\ \hline
Germany & Belgium & -1.44\\ \hline
Belgium & Netherlands & -1.37\\ \hline
United Kingdom & United States of America & -1.16\\ \hline
Germany & United States of America & -1.07\\ \hline
France & Belgium & -1.03\\ \hline
Malaysia & United Kingdom & -0.97\\ \hline
Germany & Switzerland & -0.91\\ \hline
United Kingdom & Canada & -0.90\\ \hline
Germany & United Kingdom & -0.88\\ \hline
Russia & Germany & -0.87\\ \hline
Hungary & France & -0.85\\ \hline
Zambia & United Kingdom & -0.79\\ \hline
Netherlands & United Kingdom & -0.75\\ \hline
Russia & United Kingdom & -0.73\\ \hline
Poland & Germany & -0.70\\ \hline
France & United Kingdom & -0.62\\ \hline
Finland & United Kingdom & -0.61\\ \hline
Ireland & United Kingdom & -0.59\\ \hline
France & Netherlands & -0.58\\ \hline
Malaysia & Taiwan & -0.58\\ \hline
Hungary & United States of America & -0.56\\ \hline
Finland & Germany & -0.55\\ \hline
France & United States of America & -0.53\\ \hline
Mexico & United States of America & -0.52\\ \hline
\end{tabular}
\medskip
\caption{Results of shock via link deletion for the 1965 trade web.}\label{edgetab2}
\end{table}

This type of result echoes the ``robust yet fragile" results found in network theory when subjecting networks to either targeted or random attack.  For example, networks with power laws degree distributions \cite{pnas1,sf1,sf2,Albert00} and/or small world characteristics \cite{sw1,sw2} have the property that they are very robust to random attack, but fragile in the face of targeted attack.  Our methodology shows the same type of two-pronged results -- random deletion of partnerships have little effect but there are some links which, if targeted, create substantial effects.  We emphasize, however, that the effect here is not directly linked to the degree distribution but is a product of the interaction of the network structure and the dynamics.

\begin{figure}
\begin{center}
\subfigure[Maximum partnership impact (percent of total income)]{
\includegraphics[scale=0.5]{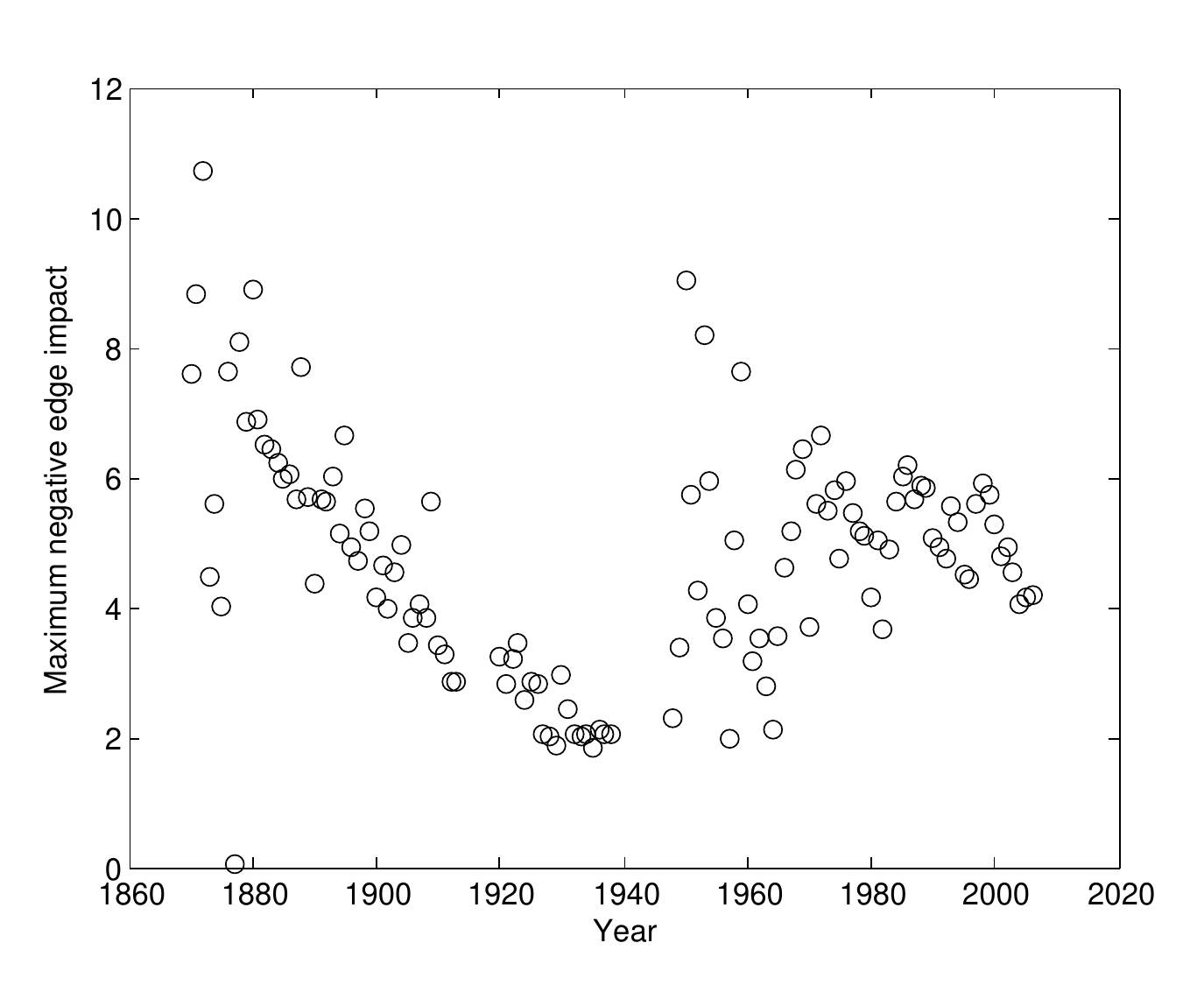}
}
\subfigure[Connectance vs. Partnership Impact]{
\includegraphics[scale=0.5]{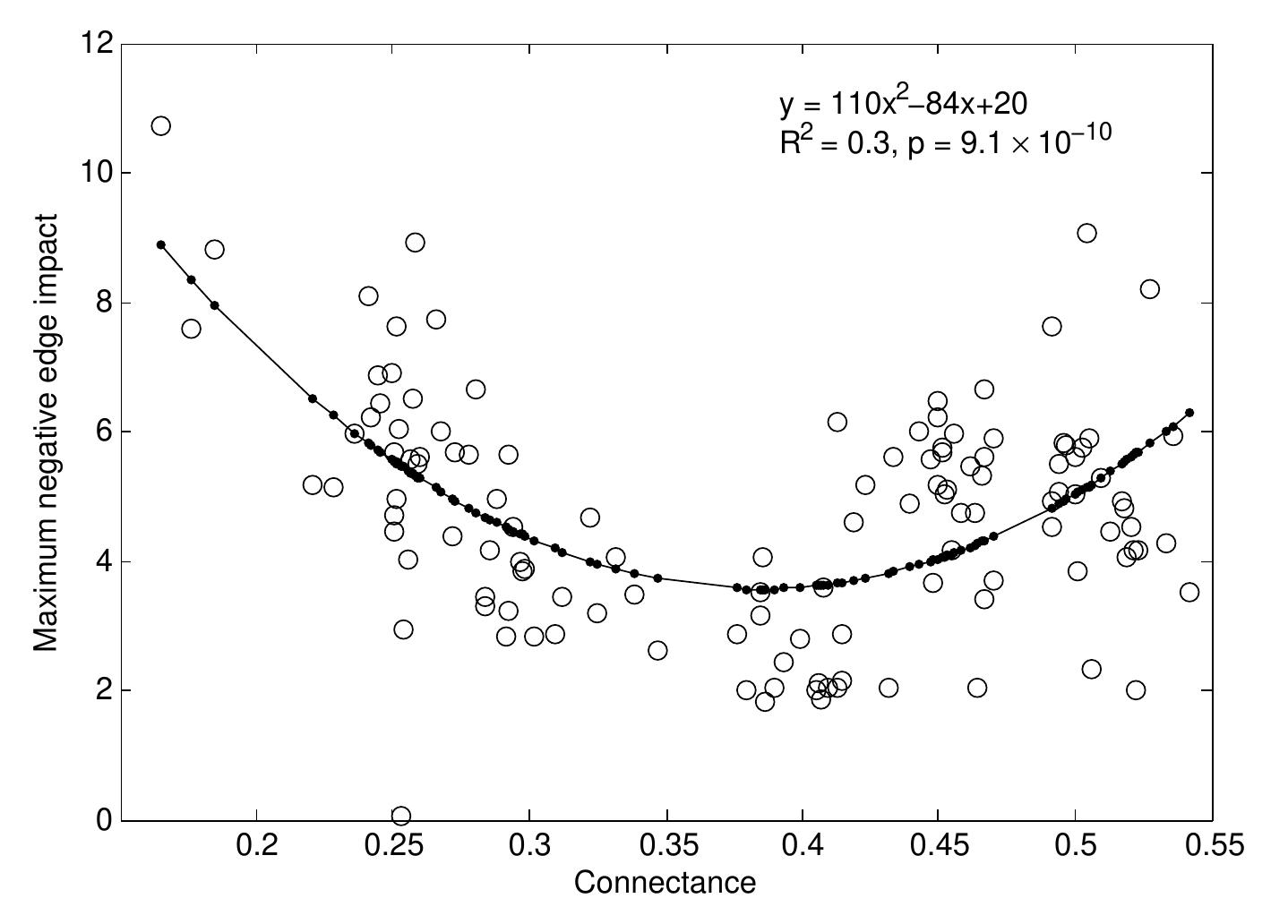}
}
\subfigure[Maximum partnership impact as multiple of total edge weight]{
\includegraphics[scale=0.5,angle=90]{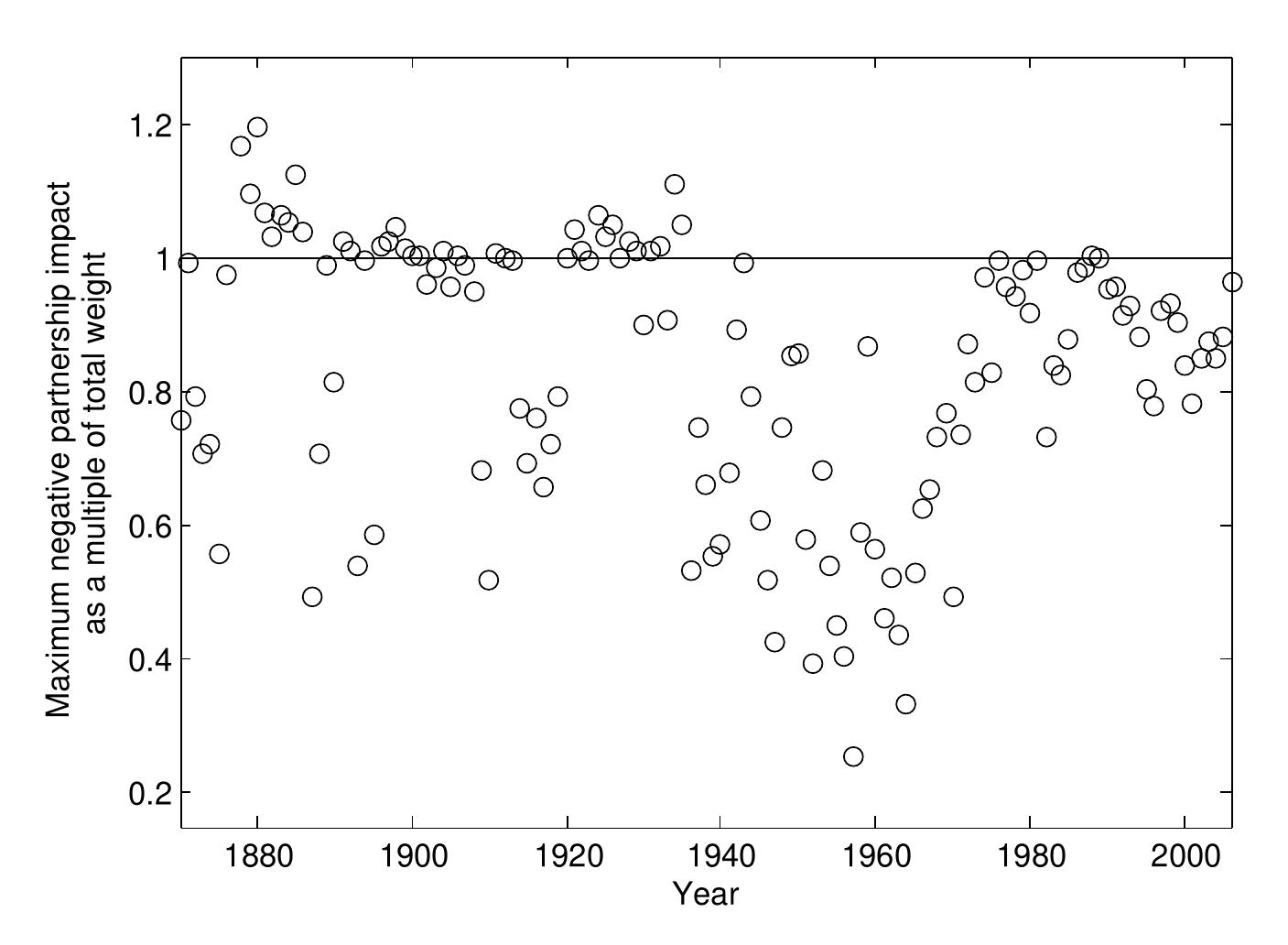}
}
\subfigure[Connectance vs. weighted edge impact]{
\includegraphics[scale=0.5]{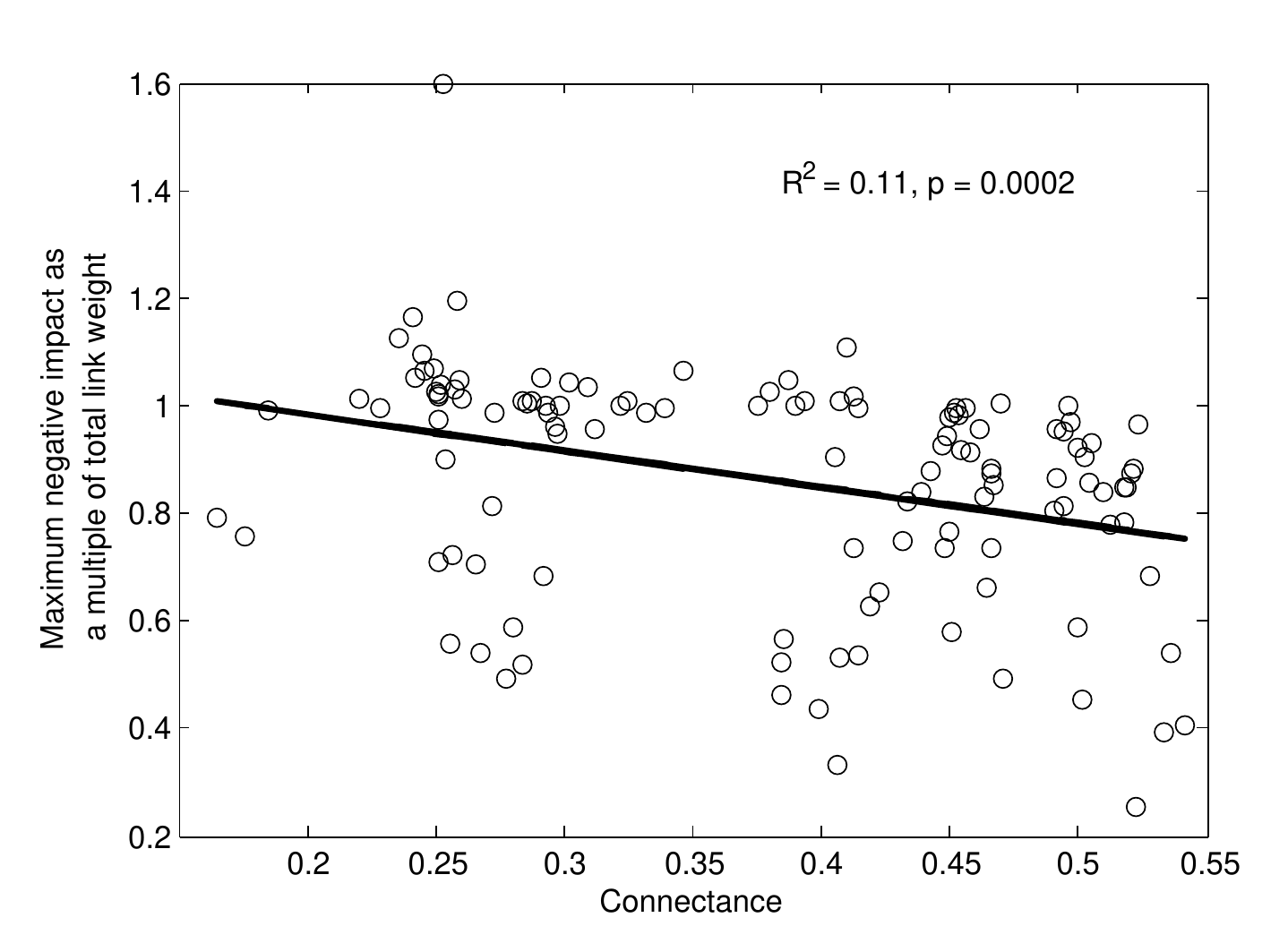}
}
\caption{Plots of the maximal impact of a single trade relationship removal, 1870--2006 (omitting the period around the World Wars).  The lefthand side shows the percentage of the total world income that is removed due to the deletion of the relationship with the most impact (top) and the same impact normalized by total edge weight (bottom). The righthand side shows a comparison of these statistics with connectance.}\label{edge-results}
\end{center}
\end{figure}

To see how this measure changes over time, we ran the same experiment on all trade webs from 1870--2006 (excluding the periods around World War I and II) and found the partnership that had the maximum negative impact.  Figure \ref{edge-results} shows the results.  On the top lefthand side, we plot the percentage of the total world income that is removed due to the deletion of the link with the most impact.  On the bottom lefthand side, we plot the percentage as a multiple of the percentage of the world income encoded in that partnership's total weight.  In the former, we see a different aspect of the same split we see in the maximal extinction analysis.  In general, there is a downward trend in the impact of removal of a single relationship.  But, after World War II, there is a period, roughly 1949--1975, where the impacts are higher than before World War II but fairly erratic.  After a transition in the 1970s, there is a resumption of the downward trend.  The righthand side provides context for the previous observations.  Before World War II, the largest impact of a single partnership was generally larger than simply removing the trade income of that partnership from the world income.  From 1949--2006, the impact is generally {\em smaller} than this removal.  The righthand side of the Figure shows the relationship between the connectance and these two measures.  We see that in the unweighted version, connectance has a quadratic relationship with the maximum partnership impact, implying that growing connectance is first correlated with decreasing impact but later with increasing impact.  The bottom righthand graph clarifies this. We see that the weighted impact is negatively correlated with connectance, i.e., as connectance grows the partnership impact decreases.  Taken together these results suggest  that connectance increases robustness in the sense of partnership impact but that its effect is mitigated by the weight associated with the trade relationship.

In the context of the results of the MEA, we see that the partnership removal analysis shows greater fragility of the network before World War II with a coherent drop and then increase after World War II.  Again, this can be plausibly explained in terms of increased international trade and trends toward globalization. Generally, with increased trade and increases in the number of trade partners - in other words, increasing connectance -- we see a general drop in the power of any given trade relationship.  In contrast, the MEA shows that increasing connectance is correlated with decreasing robustness -- deletion of all of a country's trading partners can create a substantial effect rippling throughout the network.

\section{Conclusion}
Drawing on the ideas from ecology and the analysis of food webs, we introduce the concept of extinction analysis and related techniques for the understanding of stability and robustness for economic networks.  As a first example, these tools are applied to the World Trade Web.  The analysis reveals a trend and a transition.  The trend shows a strong correlation between connectance and our robustness measures -- a negative correlation with the MEA robustness, a negative correlation with the maximum vulnerability percentage (i.e., a positive correlation with a corresponding measure of stability), and a negative correlation with edge extinction robustness.  The transition, exhibited in each case in the 1960s and 1970s, shows a rapid transition in these metrics.  In the case of the MEA, robustness sharply increases and then resumes the downward trend correlated with increasing connectance.  In the perturbation analysis, the maximum vulnerability sharply drops and then continues a decline associated with increasing connectance.  With respect to partnership deletion, we see a sharp decrease and then rebound of the maximum negative normalized impact.  We interpret this as a dramatic jump and then decline of robustness with respect to this measure.

In the context of the move towards globalization which began after World War II, we see these results as evidence of the multifaceted impact of globalization on the stability of the WTW.  We view the transitions in robustness as an indication of both the positive and negative aspects of globalization --- more trade partnerships and multiple partnerships for the same goods allow the system to recover if specific avenues of trade are removed, but, as the policy shift continues, negative implications grow.  While higher connectance provides the benefits described above, it also provides shorter paths for impacts to travel and propagate through the network.  This balance gives the network it ``robust yet fragile" characterization - a general increasing stability due to the higher connectance engendered by globalization with specific fragility to targeted shocks.

This methodology is highly adaptable and we believe it provides a useful framework for the exploration of the impact of policy changes, such as a move towards globalization, on the WTW.

\section{Acknowledgements} D.R. and N.F. were partially supported by a grant from the Alfred P. Sloan Foundation for this work. N. F. was also partially supported by the Dartmouth College Neukom Institute for Computational Science.
\bibliographystyle{plain}
\bibliography{wtw_refs}

\end{document}